# One-Parameter Scaling of the Conductivity of Si:B: A Temperature-Independent Variable-Range Hopping Prefactor.

M. P. Sarachik and P. Dai*

*Physics Department, City College of the City University of New York, New York, New York 10031*

For insulating Si:B with dopant concentrations from $0.75n_c$ to the critical concentration $n_c$, the conductivity ranging over five orders of magnitude collapses using a single scaling parameter $T^*$ onto a universal curve of the form $\sigma(T) = \sigma_0 f(T^*/T)$ with a temperature-independent prefactor of the order of Mott's minimum metallic conductivity, $\sigma_0 \approx \sigma_M \approx 0.05e^2/\hbar n_c^{-1/3}$. The function $f(T^*/T) = e^{-(T^*/T)^\beta}$ with $\beta = 1/2$ when $T^*/T > 10$, corresponding to Efros-Shklovskii variable-range hopping. For $T^*/T < 8$ the exponent $\beta = 1/3$, a value expected for Mott variable-range hopping in two rather than three dimensions. The temperature-independent prefactor implies hopping that is not mediated by phonons.

PACS numbers: 72.80.Cw, 72.80.Ng, 72.80.Sk

Hopping conduction of localized electrons in disordered insulators has been the subject of a great deal of study over the last few decades. For noninteracting electrons in disordered systems such as semiconductors, Mott [1,2] showed that a tradeoff between the exponential thermal activation due to the energy difference between the initial and final electron states on the one hand, and the exponential factor associated with the spatial overlap between the two (localized) states on the other, leads to a conductivity at low temperatures of the form:

$$\sigma \propto \sigma_M(T)e^{-(T_M/T)^\beta} \tag{1}$$

with $\beta = 1/4$ and $\beta = 1/3$ in three and two dimensions, respectively. Mott's variable range hopping (VRH) was put on a rigorous footing using a percolation formalism [3–5]. The effect of the interactions between electrons was subsequently considered by a number of theorists [6–9]. Efros and Shklovskii (ES) [8,9] showed that the long range ($1/r$) Coulomb interaction leads in an insulator to a soft gap in the one-electron density of states at the Fermi level and a depletion of low lying excitations, yielding a smaller conductivity at low temperatures of the form:

$$\sigma \propto \sigma_{ES}(T)e^{-(T_{ES}/T)^{1/2}} \tag{2}$$

with an exponent 1/2 which is independent of system dimensionality.

Various materials have been shown to obey Mott or ES variable-range hopping, and crossovers from Mott to ES hopping have been reported with decreasing temperature (when hopping energies become smaller than the gap) [10–12] and with decreasing concentration (when the gap becomes large so that the hopping electrons probe the gap) [13,14]. The expectation is that Mott variable-range hopping will always be observed near the metal-insulator transition as electron screening increases and the Coulomb gap collapses approaching the metallic phase [13,15]. Indeed, hopping exponents near 1/2 are generally found for strong electron interactions, while weak interactions (compared with hopping energies) give rise to exponents 1/4 in three dimensions and 1/3 in two dimensions. The single exception to date was reported for stressed Si:B, where the exponent was found to be 1/3 in three dimensions [16,17].

Variable-range hopping generally requires the assistance of phonons and the prefactor is theoretically expected [9] and generally found experimentally to depend on temperature. Surprisingly, there have been reports of temperature-independent prefactors near the quantum unit of conductance, $\sigma = e^2/h$ in two dimensional GaAs/$Al_xGa_{1-x}$As structures [18,19], silicon MOSFET's [20] and thin Be films [21], and approximately equal to Mott's minimum metallic value in three dimensional amorphous metal/semiconductor alloys [22] and compensated n-Ge [23]. A prefactor that does not depend on temperature implies that the hopping process is not mediated by phonons.

In this paper we report measurements of the hopping conduction of (nominally) uncompensated, insulating three-dimensional Si:B over a broad range of dopant concentrations, $0.75n_c < n < n_c$. We show that over this entire range, conductivities that vary over five orders of magnitude can be collapsed onto a common curve of the form:

$$\sigma(n, T) = \sigma_0 f[T^*/T], \tag{3}$$

using a single scaling parameter $T^*$, and a prefactor $\sigma_0$ that is independent of both temperature and dopant concentration. A similar collapse has been obtained for amorphous metal/semiconductor alloys [22]. For $T^* >> T$ (*i. e.*, low temperatures and/or low concentrations far from the metal-insulator transition) the conductivity obeys Efros-Shklovskii variable-range hopping, Eq. 2; when $T^* \leq 8T$ (higher temperatures and/or near the transition) the conductivity obeys Eq. 1 in three dimensional (unstressed) bulk Si:B, but with an exponent $\beta = 1/3$ instead of the value 1/4 expected for Mott variable-range hopping in three dimensions. The same exponent was found in experiments on Si:B where uniaxial stress instead of dopant concentration was used as the parameter to tune the material through the metal-insulator transition [16,17].

Nominally uncompensated Czochralski-grown samples of Si:B were used that contained eleven different dopant concentrations ranging from $3.24 \times 10^{18}$ to $4.30 \times 10^{18}$ $cm^{-3}$. Samples were cut into thin bars and etched in a CP-4 solution to remove any damaged surface layers. The conductivity was measured down to $\approx 50$ mK in an Oxford Model 75 dilution refrigerator by standard four-terminal techniques using low-frequency ac (15 Hz). Fig. 1 shows the conductivity of all eleven samples as a function of $T^{-1/2}$.

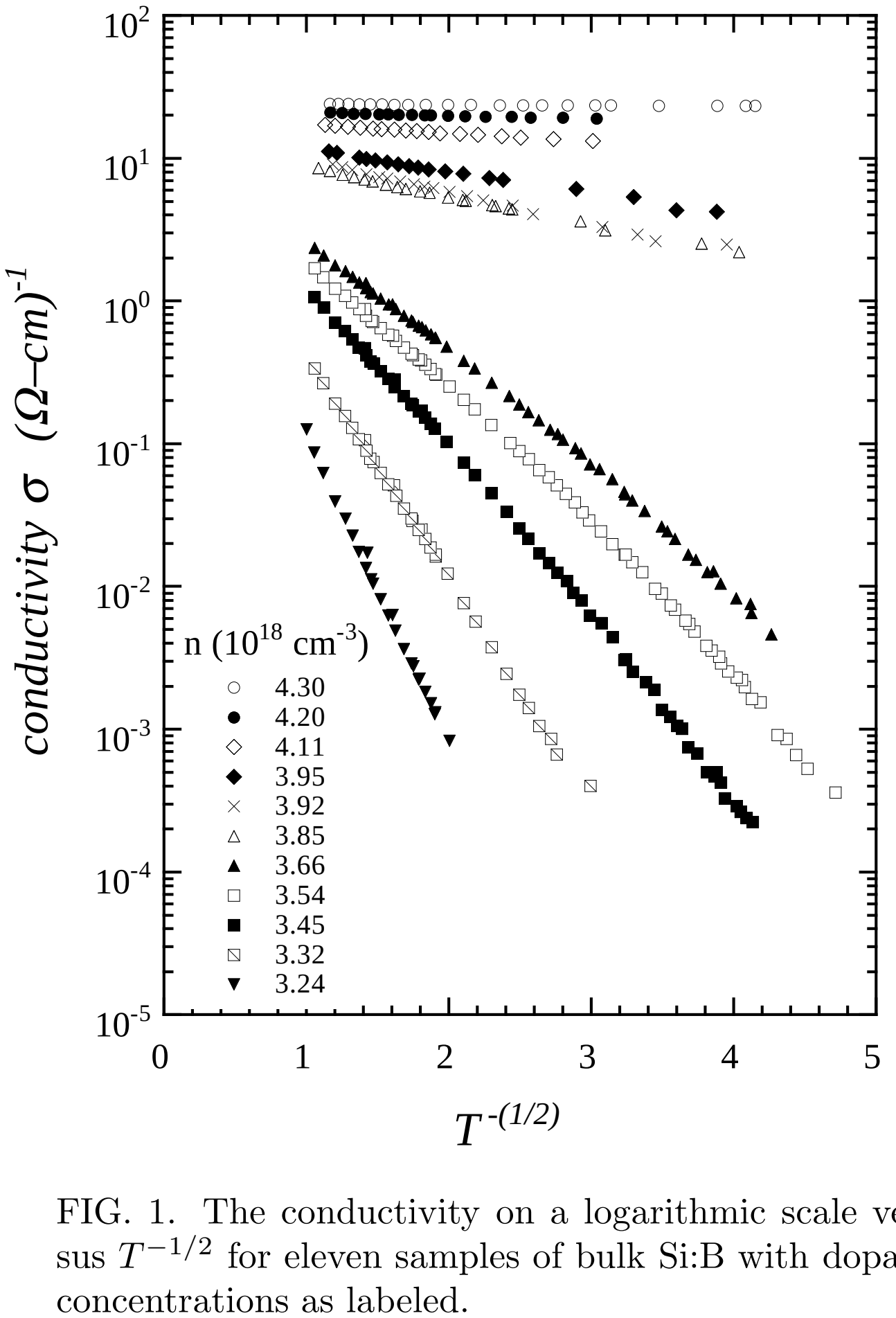

FIG. 1. The conductivity on a logarithmic scale versus $T^{-1/2}$ for eleven samples of bulk Si:B with dopant concentrations as labeled.

The data of Fig. 1 were collapsed onto a single curve by scaling the temperature by a different $T^*$ for each dopant concentration. The data collapse, shown in Fig. 2, encompasses conductivities spanning five orders of magnitude and obeys the general form, Eq. 3, with a single scaling parameter $T^*$.

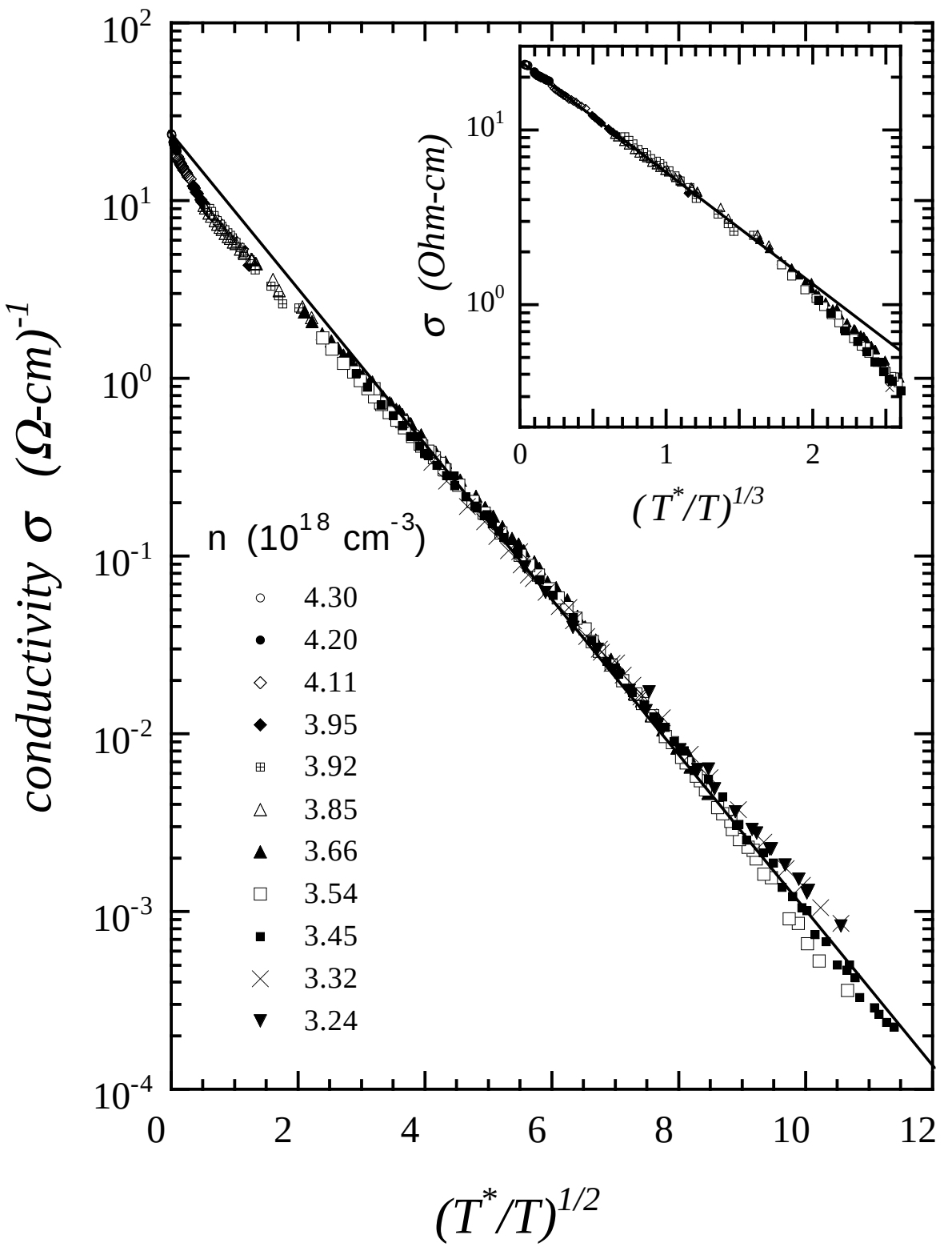

FIG. 2. Conductivity $\sigma$ of eleven samples of Si:B as a function of $T^*/T)^{1/2}$; the inset shows the conductivity as a function of $T^*/T)^{1/3}$.

For $(T^*/T)^{1/2} > 3$ (*i. e.* above $T^*/T \approx 10$), the data in the main part of Fig. 2 lie on a straight line with an extrapolated intercept $\sigma_0 = 25(\Omega\text{-cm})^{-1}$ when $((T^*/T) \rightarrow 0$. The conductivity obeys Efros-Shklovskii variable range hopping, Eq. 2, with a hopping prefactor that is independent of temperature. Deviations from this form occur at small values of $(T^*/T)^{1/2} < 3$. However, rather than Mott variable-range hopping, Eq. 1, with the exponent $\beta = 1/4$ expected in three dimensions, the conductivity instead obeys the exponentially activated form with the unexpected exponent $\beta = 1/3$. This is illustrated in the inset to Fig. 2 which shows that for $(T^*/T)^{1/3} < 2$ (*i. e.* below $T^*/T \approx 8$) the conductivity falls on a straight line when plotted as a function of $(T^*/T)^{1/3}$. In this region the conductivity is consistent with $\sigma = \sigma_0 e^{-(T^*/T)^{1/3}}$ with the same temperature-independent prefactor $\sigma_0 = 25$ (Ohm-

cm)$^{-1}$. Note that the temperature-independent prefactor is approximately Mott's minimum metallic conductivity, $\sigma_M \approx 0.05e^2/\hbar n_c^{-1/3} \approx 20$ (Ohm-cm)$^{-1}$. This is the quantum unit of conductance in three dimensions, equivalent to the quantum unit, $e^2/h$, in two dimensions.

It is noteworthy that the scaled curve of Fig. 2 includes data for three samples, $n = 4.11 \times 10^{18}$, $n = 4.20 \times 10^{18}$, and $n = 4.30 \times 10^{18}$ cm$^{-3}$, that are generally thought to be on the metallic side of the metal-insulator transition [24]. The significance of this is unclear, and requires additional careful study of the conductivity in the critical regime near the transition down to as low a temperature as possibe.

A fit to the $T^*$ used for the collapse of Fig. 2 versus concentration $n$ yields $T^* \propto (n_0 - n)^5$ with $n_0 = 4.35 \times 10^{18}$ cm$^{-3}$. Fig. 3 (a) shows the fifth power of the scaling parameter $(T^*)^{1/5}$ plotted as a function of dopant concentration, while Fig. 3 (b) shows $T^*$ versus $[(n_0 - n)/n_0]$ on a double logarithmic scale.

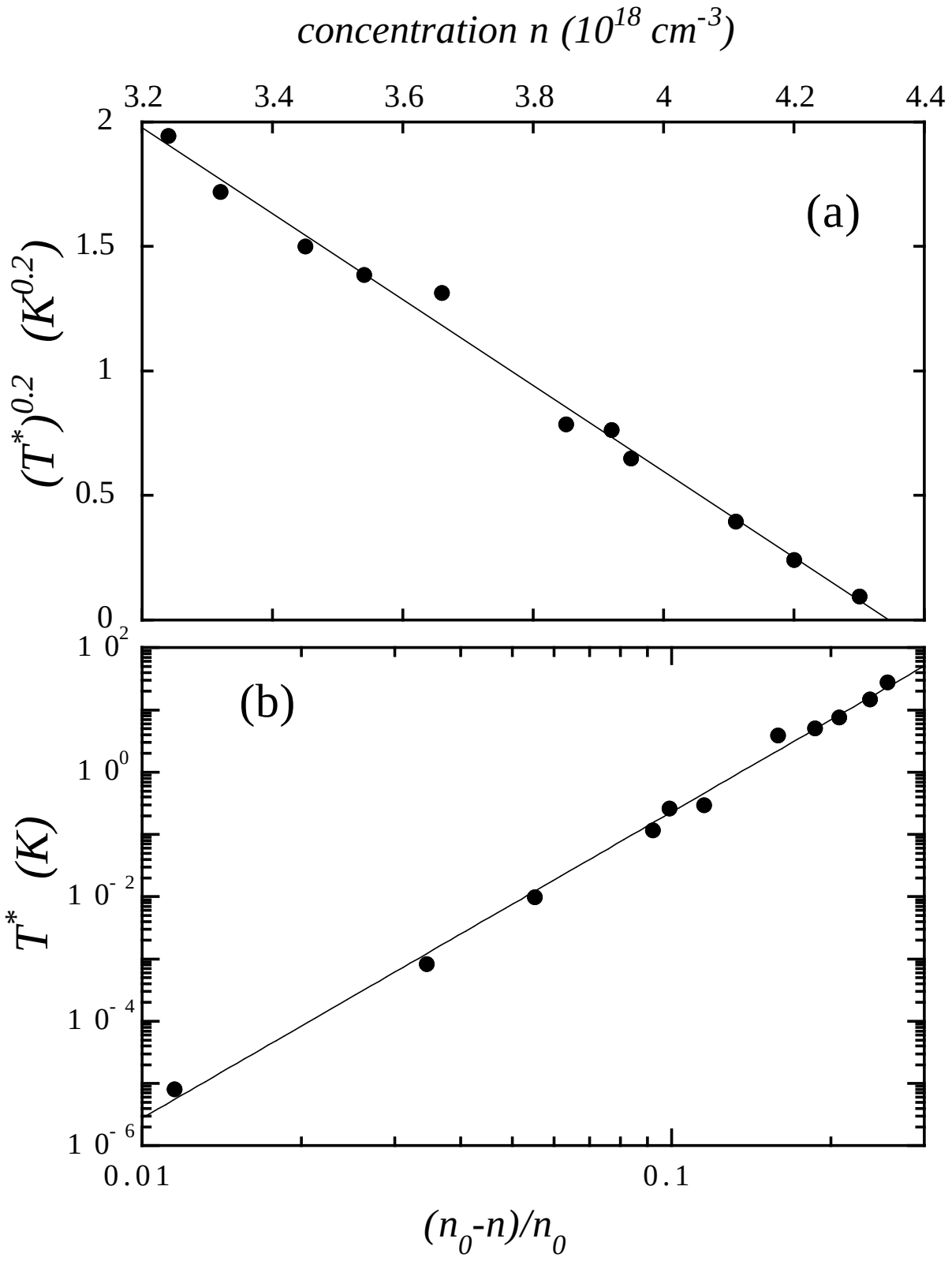

FIG. 3. (a) The fifth root of the scaling parameter $T^*$ versus dopant concentration $n$; (b) on a log-log scale, $T^*$ versus $[(n_0-n)/n_0]$ with $n_0 = 4.35\times10^{18}$ cm$^{-3}$. The absolute value of $T^*$ was chosen by setting $T^* = T_{ES}$ in the range $T^*/T > 10$ where Efros-Shklovskii variable range hopping is obeyed (see Eq. 2).

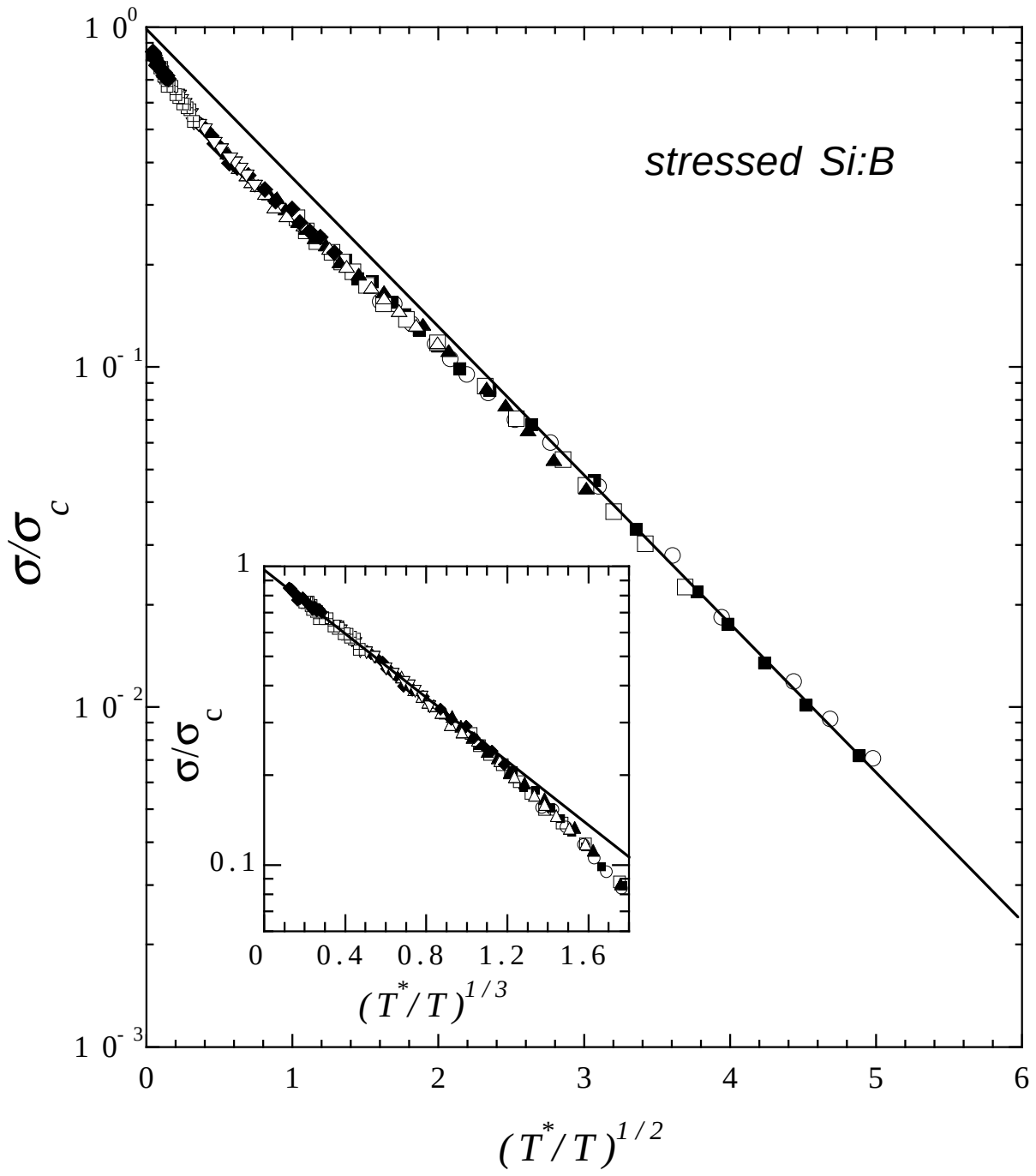

FIG. 4. For a Si:B sample subjected to uniaxial stress applied along the [001] crystalline direction, the conductivity $\sigma/\sigma_c$ is shown as a function of $(T^*/T)^{1/2}$; the critical conductivity $\sigma_c \propto T^{1/2}$. The uniaxial stress was varied between the critical value $S_c \approx 613$ bar and $S = 873$ bar $\approx 1.4 n_c$ [16,17]. The inset shows $\sigma/\sigma_c$ versus $(T^*/T)^{1/3}$.

It is interesting to compare the collapsed data of Fig. 2 obtained for samples with different dopant concentrations, $n$, with similar data measured for the same material where uniaxial stress, $S$, was used instead as the tuning parameter. As shown in Fig. 4, a very similar data collapse was obtained: Efros-Shklovskii variable-range hopping for large values of $(T^*/T)$ and exponentially activated variable-range hopping with an unexpected exponent $\beta = 1/3$ for small $(T^*/T)$ . However, there is an important difference between the two cases: when stress is used as the tuning parameter, it is $\sigma/\sigma_c$ rather than $\sigma$ itself that exhibits the behavior shown in Fig. 4. The critical temperature dependence of the conductivity for stressed Si:B was found to be $\sigma_c \propto T^{1/2}$, so that $\sigma \propto T^{1/2} e^{-(T^*/T)^{1/2}}$. The hopping prefactor, $\sigma_0 = aT^{1/2}$, is thus *not* independent of temperature when stress is applied. This enigmatic difference requires further investigation. Nevertheless, in both stressed and unstressed Si:B, the temperature-dependence near the metal-insulator transition ($(T^*/T) = 0$ at $n_c$) is expo-nentially activated with exponent $\beta = 1/3$; in neither case is an acceptable fit obtained for $\beta = 1/4$, the value expected in three dimensions. This conclusion is based on data for samples with scaled conductivities that vary by more than an order of magnitude (see inset to Fig. 2).

To summarize, for the uncompensated doped semiconductor Si:B, we have obtained a full collapse on the insulating side of the metal-insulator transition of the conductivity spanning five orders of magnitude using a single scaling parameter $T^*$. The conductivity is found to obey the exponentionally activated variable-range hopping form, $\sigma = \sigma_0 \exp^{[-(T^*/T)^\beta]}$ with a prefactor $\sigma_0$ that does not depend on temperature or concentration. There have been several other claims of temperature-independent prefactors approximately equal to $e^2/h$ in two-dimensional systems [18–21] and equal to Mott's minimum metallic conductivity in three-dimensional systems [22,23]. A prefactor that does not depend on temperature implies that the hopping in these systems is not mediated by phonons. The possibility of variable-range hopping involving electron-electron interactions (instead of the usual electron-phonon mechanism) has been discussed by a number of authors [25–28], but is not well understood. Most recently, Kozub, Baranovskii and Shlimak [28] have suggested that hopping takes place by resonant tunneling between transport states that fluctuate in energy. What conditions must be met in order that the prefactor be independent of temperature, and how the hopping proceeds in such cases, are very interesting open questions.

An additional finding is that the conductivity of uncompensated Si:B is exponentionally activated with a variable-range hopping exponent $\beta = 1/3$ at temperatures and dopant densities where one expects Mott variable-range hopping with exponent $\beta = 1/4$ in three dimensions. It is of interest to determine whether similar unexpected behavior obtains in other three-dimensional systems, and what mechanism is responsible for it.

We thank A. Aharony, R. N. Bhatt, M. Raikh, S. Baranovksii, A. L. Efros and B. I. Shklovskii for useful discussions and comments on the manuscript. This work was supported by U. S. Department of Energy grant No. DE-FG02-84ER45153.

---

* Present address: Oplink Communication Inc., San Jose, CA 95134-1801; peterd@oplink.com.